\documentclass[twoside,nohyper]{JHEP3}
\usepackage{amsmath,amssymb,longtable,dsfont}

\newcommand{\ve}{\varepsilon}

\newcommand{\beq}{\begin{equation}}
\newcommand{\eeq}{\end{equation}}
\newcommand{\bea}{\begin{eqnarray}}
\newcommand{\eea}{\end{eqnarray}}
\newcommand{\bes}{\begin{split}}
\newcommand{\ees}{\end{split}}
\newcommand{\bed}{\begin{displaymath}}
\newcommand{\eed}{\end{displaymath}}

\newcommand{\ot}{{\cal O}(q^3)}

\newcommand{\la}{\langle}
\newcommand{\ra}{\rangle}
\newcommand{\nn}{\nonumber}

\title{
Meson-Baryon Effective Chiral Lagrangians 
at ${\cal O}(q^3)$: Erratum}
\author{Jos\'e Antonio Oller and Michela Verbeni\\
Departamento de F\'\i sica, 
Universidad de Murcia, E-30071 Murcia, Spain.\\
E-mail: \email{oller@um.es} , \email{mverbeni@ugr.es}}
\author{Joaquim Prades\\
CAFPE and Departamento de F\'\i sica Te\'orica y del Cosmos,
Universidad de Granada, Campus de Fuente Nueva, E-18002 Granada, Spain\\
 and 
Theory Unit, Physics Department, CERN, CH-1211 Gen\`eve 23,
Switzerland.\\
E-mail: \email{Prades@ugr.es}}

\abstract{After our work \cite{OVP06}
was published, Frink and Mei{\ss}ner \cite{FM06}
pointed out that the ${\cal O}(q^3)$  three-flavour  meson-baryon 
chiral Lagrangian presented there was not minimal.
 Here, we discuss their findings and revise ours
accordingly. We find out  eight monomials in the Lagrangian
presented in \cite{OVP06}
 are not independent, but in addition,  two monomials were
wrongly discarded there which, as a result, makes the agreement
with \cite{FM06} in the number of  independent monomials complete.} 

\preprint{CERN-PH-TH/2007-004  \\January 2007}

\begin{document}

\def\query#1{\marginpar{\begin{flushleft}\footnotesize#1\end{flushleft}}}


 Recently, Frink and 
Mei{\ss}ner \cite{FM06} pointed out that one can further reduce 
the number of monomials present in the ${\cal O}(q^3)$ Lagrangian 
of \cite{OVP06}  by six, passing from 84 in \cite{OVP06} to 
78 in \cite{FM06}.  
Here, we revise
our Lagrangian and discuss also the findings of \cite{FM06}
 since  some of them are not accurate and require clarification.
 We find out that actually one can reduce by eight the number
of independent monomials in \cite{OVP06}  but in addition,  two monomials
were wrongly discarded in \cite{OVP06} which, as a result,
makes the agreement in the number of independent monomials
with \cite{FM06}  complete.
We refer to \cite{OVP06} for the presentation 
 of the  building blocks and techniques 
 employed in the construction of the monomials, where it is discussed 
 in detail.

 Some Cayley-Hamilton relations involving monomials with five flavour  
matrices  were missed  by us, 
as correctly noticed in \cite{FM06}. The technicalities 
 of this point  are explained in detail in the 
 Appendix A of \cite{FM06}. 
Along these lines, we find three Cayley-Hamilton relations among
the monomials $O_{12}$ to $O_{25}$ of our Lagrangian that were not taken
into account there. If these Cayley-Hamilton relations 
are used to discard only monomials involving the product of 
two flavour traces, then one monomial between $O_{20}$, $O_{22}$ 
and $O_{24}$ and two more monomials between $O_{21}$, $O_{23}$ and $O_{25}$ 
can be neglected.
We choose to cast away $O_{22}$, $O_{23}$ and $O_{25}$.
Thus, we agree with \cite{FM06}
that Cayley-Hamilton relations can be used to further reject
three monomials from $O_{12}$ to $O_{25}$ in our Lagrangian.  
However, it is not possible to simultaneously 
disregard the monomials  $O_{20}$, $O_{21}$ and $O_{22}$ 
from that basis.

 We find two other Cayley-Hamilton relations among 
the monomials $O_{31}$ to $O_{37}$ in our Lagrangian not considered
before. They allow to discard 
two  monomials between $O_{35}$, $O_{36}$ and $O_{37}$, 
as already remarked in \cite{FM06}.
We choose to cast aside $O_{35}$ and $O_{36}$.

 Another Cayley-Hamilton relation among  the monomials
$O_{38}$ to $O_{43}$ in our Lagrangian, not used in ref.\cite{OVP06}, 
is found now.
  This fact  is not 
commented in \cite{FM06}.  In this way, one can remove
another monomial that we choose to be $O_{43}$.
 
 In \cite{OVP06} we used  a Cayley-Hamilton relation to cast away
 the one flavour trace monomial,
\beq
\widehat O_{35} = 
i\left(\la\bar{B} \{u^\nu,u^\rho\} \sigma^{\lambda\tau} D_\rho B 
 u^\mu \ra - \la\bar{B}\overleftarrow{D}_\rho \{u^\nu,u^\rho\}
\sigma^{\lambda\tau} B u^\mu \ra\right)\ve_{\mu\nu\lambda\tau}~, 
\eeq
while all the other monomials neglected because of using 
the Cayley-Hamilton theorem 
contained more than one flavour trace. Here,  due to  large
$N_c$ counting, we prefer to neglect the two trace monomial
$O_{42}$ in \cite{OVP06} and put back $\widehat O_{33}$ in our new basis 
for the ${\cal O}(q^3)$ Lagrangian.

Apart from the missed Cayley-Hamilton relations in 
our Lagrangian, Frink and Mei{\ss}ner \cite{FM06} also realized that 
only the symmetric combination of $O_9$ 
and $O_{10}$ in \cite{OVP06} is independent. Hence, only one of these 
two monomials should be considered  and we keep $O_9$. 
Since we found difficulties in understanding the  
argumentation given in \cite{FM06}, we reproduce 
here our way of deriving such  relationship between  $O_9$ and 
$O_{10}$. We proceed as follows. Taking into account that
\beq 
D_\nu u_\rho-D_\rho u_\nu=f^-_{\rho\nu}~,
\label{2.10}
\eeq
 see eq.(2.10) of \cite{OVP06},
 the difference between   
$O_9=i\la \bar{B}u^\mu \sigma_{\mu \nu}D_\rho B\,h^{\nu\rho}\ra-i\la 
\bar{B}\overleftarrow{D}_\rho u^\mu \sigma_{\mu\nu}B h^{\nu\rho}\ra$, and 
$i\la \bar{B}u^\mu \sigma_{\mu \nu}D_\rho B\,D^\nu u^\rho\ra-i\la 
\bar{B}\overleftarrow{D}_\rho u^\mu \sigma_{\mu\nu}B D^\nu u^\rho\ra$, 
is accounted for by the monomial $O_{82}$ of ref.\cite{OVP06}, 
or by our present ${\widehat O}_{76}$ of Table \ref{table}. 
Then, neglecting  a global divergence, 
\bea
O_9 \to &-&i\la D^\nu \bar{B} u^\mu \sigma_{\mu \nu} D_\rho B\, u^\rho\ra
-i\la \bar{B} D^\nu u^\mu\sigma_{\mu\nu}D_\rho B\,u^\rho\ra
-i \la \bar{B} u^\mu \sigma_{\mu\nu} D^\nu D_\rho B\, u^\rho\ra\nn\\
&+&i\la D^\nu D_\rho \bar{B} u^\mu \sigma_{\mu \nu} B\, u^\rho\ra
+i\la D_\rho \bar{B} D^\nu u^\mu \sigma_{\mu \nu} B\, u^\rho\ra
+i\la D_\rho \bar{B} u^\mu \sigma_{\mu\nu} D^\nu B\, u^\rho\ra~,
\label{justeq}\eea 
where other monomials already accounted for are not written and 
this is why we use the right pointing arrow.
The second term on each of the lines of eq.(\ref{justeq})  can be written
again in terms of monomials with $f^-_{\mu\nu}$ because of eq.(\ref{2.10}),
 since $D^\nu u^\mu$ is contracted with the 
antisymmetric tensor $\sigma_{\nu\mu}$.  
The resulting structures are taken into account by the monomial 
$\widehat O_{75}$ in Table \ref{table}. 
In this way we are left with
\bea
O_9&\rightarrow& -i\la D^\nu \bar{B} u^\mu 
 \sigma_{\mu\nu} D^\rho B\, u_\rho\ra
-i\la \bar{B} u^\mu \sigma_{\mu\nu}
D^\nu D^\rho B \,u_\rho\ra \nn\\
&&+i\la D^\rho \bar{B} u^\mu \sigma_{\mu \nu} 
D^\nu B \,u_\rho\ra
+i\la D^\nu D^\rho \bar{B} u^\mu \sigma_{\mu \nu} B\, 
u_\rho\ra~.
\eea
Employing the relation 
$-i\sigma_{\mu\nu}=g_{\mu\nu}-\gamma_\nu\gamma_\mu $ in the first 
and fourth  monomials above and 
$+i\sigma_{\mu\nu}=g_{\mu\nu}-\gamma_\mu\gamma_\nu$ in the 
second and third ones, one can write
\bea
O_9&\rightarrow&-4\la \bar{B}u^\nu D_\nu D_\rho B\,u^\rho\ra-2\la 
\bar{B} u^\nu D_\rho B\, D_\nu u^\rho\ra
-2\la \bar{B} D_\rho u^\nu D_\nu B\,u^\rho\ra~,
\label{o9new}  
\eea
where the equation of motion of baryons has been used to remove 
those terms involving 
$\gamma^\nu D_\nu B$ and $D_\nu \bar{B} \gamma^\nu$, see eq.(4.2) of
\cite{FM06}. One can proceed analogously for $O_{10}$ and 
then exactly the same combination of monomials as in (\ref{o9new}) 
is found. 
Hence, only the symmetric combination of $O_9$ and $O_{10}$ is 
independent, while the difference can be written in terms of 
other monomials already taken into account.

Frink and Mei{\ss}ner also noticed that the 
index ordering in the monomials
$O_{31}$, $O_{33}$ and $O_{34}$ in \cite{OVP06} do not match the conditions
imposed by charge conjugation invariance.  We want to point out that the
difference between the index ordering in \cite{OVP06}
and that which is exactly invariant under charge conjugation 
is ${\cal O}(q^4)$. However, we prefer monomials
in the Lagrangian  which are exactly charge conjugation invariant, 
because charge conjugation is a symmetry of strong interactions
--see our comments in \cite{OVP06}.  Then, we now take the 
ordering in the indices such that these monomials are exactly charge
conjugation invariant.

 As pointed out in \cite{FM06} the relative sign between 
the two flavour traces  in $O_{41}$ should be plus instead of the 
minus in \cite{OVP06}.  
 Once this is corrected $O_{41}$ becomes of ${\cal O}(q^4)$. 
Then, the comment   at the end of Section 5 of \cite{OVP06},
  though correct,  is not relevant. 
  
   In addition, we notice that two independent monomials were
wrongly discarded in \cite{OVP06}.  These  monomials are
\beq
\widehat O_{32} = 
\la\bar{B} \left[\left[
u_\mu, u_\nu \right], u^\rho \right] \gamma_5 \sigma^{\mu\nu}
D_{\rho}  B\ra - \la\bar{B}\overleftarrow{D}_{\rho} 
\left[ \left[u_\mu, u_\nu \right], u^\rho \right]
 \gamma_5\sigma^{\mu\nu} B\ra \, 
\eeq
 and
\beq
\widehat O_{33} = 
\la\bar{B} \gamma_5 \sigma^{\mu\nu} D_{\rho}  B
\left[\left[u_\mu, u_\nu \right], u^\rho \right] \ra -
 \la\bar{B}\overleftarrow{D}_{\rho}
 \gamma_5\sigma^{\mu\nu} B
 \left[ \left[u_\mu, u_\nu \right], u^\rho \right] \ra \, .
\eeq

Summarizing  the discussion above, we can  take away from
 our  ${\cal O}(q^3)$  three-flavour  meson-baryon Lagrangian  
 the following eight monomials:
$O_{10}$, $O_{22}$, $O_{23}$, $O_{25}$, $O_{35}$, $O_{36}$,
$O_{41}$ and  $O_{43}$. In addition, 
we exchange $O_{42}$ by  $\widehat O_{35}$ and  add two
monomials, namely, $\widehat O_{32}$ and $\widehat O_{33}$, 
 not included in \cite{OVP06}.
 We therefore end up with 78 independent monomials  in the
$SU(3)$ meson-baryon chiral  Lagrangian at ${\cal O}(q^3)$
and agree fully with  \cite{FM06}.   
   We give the complete list of the 
monomials present in the minimal $SU(3)$ meson-baryon
chiral invariant Lagrangian in Table \ref{table}. 
\beq
{\cal L}_{MB}^{(3)} = \sum_{i=1}^{78} \, h_i \,  {\widehat O}_i \, .
\eeq

\renewcommand{\arraystretch}{1.5}
\begin{longtable}[r]{|c|c|c|}
\hline
$i$&$\widehat O_i$ & Contributes to vertex\\
\hline
\hline
\endhead
\hline
\caption[]{\rule{0cm}{2em}}
\endfoot
\hline
\caption[]
{\label{table} Minimal set of linearly independent  monomials
of the  $SU(3)$ chiral meson-baryon Lagrangian of $\ot$. On the third
column we give  the vertex  with the minimal number of mesons and photons
to which each term contributes.}
\endlastfoot
1&$i\left(\la\bar{B}\gamma_\mu
D_{\nu\rho}B[u^\mu,h^{\nu\rho}]\ra + \la\bar{B}\overleftarrow{D}_{\nu\rho}
\gamma_\mu B[u^\mu,h^{\nu\rho}]\ra\right)$& $M_1 B_1 \to M_2 B_2$\\ 
2&$i\left(\la\bar{B}[u^\mu,h^{\nu\rho}]\gamma_\mu
D_{\nu\rho}B\ra + \la\bar{B}\overleftarrow{D}_{\nu\rho}
[u^\mu,h^{\nu\rho}]\gamma_\mu B\ra\right)$ & $M_1 B_1 \to M_2 B_2$\\ 
3&$i\left(\la\bar{B}u^\mu\ra\la h^{\nu\rho} \gamma_\mu
    D_{\nu\rho}B\ra - \la\bar{B}\overleftarrow{D}_{\nu\rho} h^{\nu\rho}\ra
\la u^\mu\gamma_\mu B\ra\right)$& $M_1 B_1 \to M_2 B_2$\\
4&$i\la\bar{B}[u_\mu,h^{\mu\nu}]\gamma_\nu B\ra$& $M_1 B_1 \to M_2 B_2$\\
5&$i\la\bar{B}\gamma_\nu B[u_\mu,h^{\mu\nu}]\ra$& $M_1 B_1 \to M_2B_2$\\
6&$i\left(\la\bar{B}u_\mu\ra\la h^{\mu\nu}\gamma_\nu B\ra
-\la\bar{B}h^{\mu\nu}\ra\la u_\mu\gamma_\nu B\ra\right)$& $M_1 B_1 \to M_2B_2$\\
7&$i\la\bar{B}\sigma_{\mu\nu} D_\rho
    B\{u^\mu,h^{\nu\rho}\}\ra-i\la\bar{B}\overleftarrow{D}_\rho\sigma_{\mu\nu} 
B \{u^\mu,h^{\nu\rho}\}\ra$& $M_1 B_1 \to M_2B_2$\\
8&$i\la\bar{B}\{u^\mu,h^{\nu\rho}\}\sigma_{\mu\nu} D_\rho
    B\ra-i\la\bar{B}\overleftarrow{D}_\rho\{u^\mu,h^{\nu\rho}\}
\sigma_{\mu\nu}B\ra$& $M_1 B_1 \to M_2B_2$\\
9&$i\la\bar{B}u^\mu \sigma_{\mu\nu} D_\rho B h^{\nu\rho}\ra
-i\la\bar{B}\overleftarrow{D}_\rho u^\mu \sigma_{\mu\nu} B h^{\nu\rho}\ra$
& $M_1 B_1 \to M_2B_2$\\
10&$i\left(\la\bar{B}\sigma_{\mu\nu} D_\rho B\ra
-\la\bar{B}\overleftarrow{D}_\rho\sigma_{\mu\nu}B\ra\right)
\la u^\mu h^{\nu\rho}\ra$& $M_1 B_1 \to M_2B_2$\\
11&$\la\bar{B}\gamma_5\gamma_\nu B\{u_\mu u^\mu,u^\nu\}\ra$&$M_1 B_1
\to M_2 M_3 B_2$\\
12&$\la\bar{B}\gamma_5\gamma_\nu B u_\mu u^\nu u^\mu\ra$&$M_1 B_1
\to M_2 M_3 B_2$\\
13&$\la\bar{B}u_\mu\gamma_5\gamma_\nu B\{u^\mu,u^\nu\}\ra$&$M_1 B_1
\to M_2 M_3 B_2$\\
14&$\la\bar{B}u_\mu u^\mu\gamma_5\gamma_\nu B u^\nu\ra$&$M_1 B_1
\to M_2 M_3 B_2$\\
15&$\la\bar{B}\{u_\mu u^\mu,u^\nu\}\gamma_5\gamma_\nu B\ra$&$M_1 B_1
\to M_2 M_3 B_2$\\
16&$\la\bar{B}\{u^\mu,u^\nu\}\gamma_5\gamma_\nu B u_\mu\ra$&$M_1 B_1
\to M_2 M_3 B_2$\\
17&$\la\bar{B}u_\mu u^\nu u^\mu\gamma_5\gamma_\nu B\ra$&$M_1 B_1
\to M_2 M_3 B_2$\\
18&$\la\bar{B}u^\nu\gamma_5\gamma_\nu B u_\mu u^\mu\ra$&$M_1 B_1
\to M_2 M_3 B_2$\\
19&$\la\bar{B}\{u^\nu,\gamma_5\gamma_\nu B\}\ra\la u_\mu u^\mu\ra$&$M_1 B_1
\to M_2 M_3 B_2$\\
20&$\la\bar{B}[u^\nu,\gamma_5\gamma_\nu B]\ra\la u_\mu u^\mu\ra$&$M_1 B_1
\to M_2 M_3 B_2$\\
21&$\la\bar{B}\gamma_5\gamma_\nu B\ra\la u_\mu u^\mu u^\nu\ra$&$M_1 B_1
\to M_2 M_3 B_2$\\
22&$i\la\bar{B}\gamma^\tau
  B\{[u^\mu,u^\nu],u^\rho\}\ra\ve_{\mu\nu\rho\tau}$&$M_1 B_1
\to M_2 M_3 B_2$\\
23&$i\la\bar{B}\{[u^\mu,u^\nu],u^\rho\}\gamma^\tau
  B\ra \ve_{\mu\nu\rho\tau}$&$M_1 B_1
\to M_2 M_3 B_2$\\
24&$i\la\bar{B}[u^\mu,u^\nu]\gamma^\tau B u^\rho\ra
  \ve_{\mu\nu\rho\tau}$&$M_1 B_1
\to M_2 M_3 B_2$\\
25& $i\la\bar{B}u^\rho\gamma^\tau B[u^\mu,u^\nu]\ra
  \ve_{\mu\nu\rho\tau}$&$M_1 B_1
\to M_2 M_3 B_2$\\
26&$i\la\bar{B}\gamma^\tau
  B\ra\la[u^\mu,u^\nu]u^\rho\ra\ve_{\mu\nu\rho\tau}$&$M_1 B_1
\to M_2 M_3 B_2$\\
27&$\la\bar{B}\gamma_5\gamma_\mu D_{\nu\rho}B u^\mu u^\nu u^\rho\ra
+ \la\bar{B}\overleftarrow{D}_{\nu\rho} \gamma_5\gamma_\mu B
u^\rho u^\nu u^\mu\ra$&$M_1 B_1
\to M_2 M_3 B_2$\\
28&$\la\bar{B}u^\mu\gamma_5\gamma_\mu D_{\nu\rho}Bu^\nu u^\rho\ra
+ \la\bar{B}\overleftarrow{D}_{\nu\rho} u^\mu \gamma_5\gamma_\mu B
u^\rho u^\nu\ra$&$M_1 B_1
\to M_2 M_3 B_2$\\
29&$\la\bar{B}u^\mu u^\nu\gamma_5\gamma_\mu D_{\nu\rho}B u^\rho\ra
+\la\bar{B}\overleftarrow{D}_{\nu\rho} u^\nu u^\mu \gamma_5\gamma_\mu
B u^\rho\ra$&$M_1 B_1\to M_2 M_3 B_2$\\
30&$\la\bar{B}u^\mu u^\nu u^\rho \gamma_5\gamma_\mu D_{\nu\rho}
  B\ra + \la\bar{B}\overleftarrow{D}_{\nu\rho} u^\rho u^\nu u^\mu
 \gamma_5\gamma_\mu B\ra$&$M_1 B_1\to M_2 M_3 B_2$\\
31&$\left(\la\bar{B}\gamma_5\gamma_\mu D_{\nu\rho}B\ra
+ \la\bar{B}\overleftarrow{D}_{\nu\rho}\gamma_5\gamma_\mu B\ra\right)
\la u^\mu u^\nu u^\rho\ra$&$M_1 B_1\to M_2 M_3 B_2$\\
32&$\la\bar{B} \left[\left[
u_\mu, u_\nu \right], u^\rho \right] \gamma_5 \sigma^{\mu\nu}
D_{\rho}  B\ra - \la\bar{B}\overleftarrow{D}_{\rho} 
\left[ \left[u_\mu, u_\nu \right], u^\rho \right]
 \gamma_5\sigma^{\mu\nu} B\ra$&$M_1 B_1\to M_2 M_3 B_2$\\
33&$
\la\bar{B} \gamma_5 \sigma^{\mu\nu} D_{\rho}  B
\left[\left[u_\mu, u_\nu \right], u^\rho \right] \ra -
 \la\bar{B}\overleftarrow{D}_{\rho}
 \gamma_5\sigma^{\mu\nu} B
 \left[ \left[u_\mu, u_\nu \right], u^\rho \right] \ra 
$&$M_1 B_1\to M_2 M_3 B_2$\\
34&$i\left(\la\bar{B}u^\mu\sigma^{\lambda\tau} D_\rho
  B\{u^\nu,u^\rho\}\ra - \la\bar{B}\overleftarrow{D}_\rho u^\mu
  \sigma^{\lambda\tau} B \{u^\nu,u^\rho\}\ra\right)\ve_{\mu\nu\lambda\tau}$
&$M_1 B_1\to M_2 M_3 B_2$\\
35&$i\left(\la\bar{B} \{u^\nu,u^\rho\} \sigma^{\lambda\tau} D_\rho B 
 u^\mu \ra - \la\bar{B}\overleftarrow{D}_\rho \{u^\nu,u^\rho\}
\sigma^{\lambda\tau} B u^\mu \ra\right)\ve_{\mu\nu\lambda\tau}$
&$M_1 B_1\to M_2 M_3 B_2$\\
36&$i\left(\la\bar{B}\{u^\mu,\sigma^{\lambda\tau} D_\rho B\}\ra
-\la\bar{B}\overleftarrow{D}_\rho\{u^\mu,\sigma^{\lambda\tau}B\}\ra\right)
\la u^\nu u^\rho\ra \ve_{\mu\nu\lambda\tau}$&$M_1 B_1\to M_2 M_3 B_2$\\
37&$i\left(\la\bar{B}[u^\mu,\sigma^{\lambda\tau} D_\rho B]\ra
-\la\bar{B}\overleftarrow{D}_\rho[u^\mu,\sigma^{\lambda\tau}B]\ra\right)
\la u^\nu u^\rho\ra \ve_{\mu\nu\lambda\tau}$&$M_1 B_1\to M_2 M_3 B_2$\\
38&$\la\bar{B}u^\mu \gamma_5\gamma_\mu B \chi_+\ra$& $B_1 \to M_1 B_2$\\
39&$\la\bar{B}\chi_+ \gamma_5\gamma_\mu B u^\mu\ra$& $B_1 \to M_1 B_2$\\
40&$\la\bar{B}u^\mu \gamma_5\gamma_\mu B\ra\la\chi_+\ra$& $B_1 \to M_1 B_2$\\
41&$\la\bar{B}\gamma_5\gamma_\mu B u^\mu\ra\la\chi_+\ra$& $B_1 \to M_1 B_2$\\
42&$\la\bar{B}\gamma_5\gamma_\mu B\ra\la u^\mu \chi_+\ra$& $B_1 \to M_1 B_2$\\
43&$\la\bar{B}\gamma_5\gamma_\mu B\{u^\mu,\chi_+\}\ra$& $B_1 \to M_1 B_2$\\
44&$\la\bar{B}\{u^\mu,\chi_+\}\gamma_5\gamma_\mu B\ra$& $B_1 \to M_1 B_2$\\
45&$\la\bar{B}\{\chi_-,\gamma_5 B\}\ra$& $B_1 \to M_1 B_2$\\
46&$\la\bar{B}[\chi_-,\gamma_5 B]\ra$ &$B_1 \to M_1 B_2$\\
47&$\la\bar{B}\gamma_5 B\ra\la\chi_-\ra$&$B_1 \to M_1 B_2$\\
48&$\la\bar{B}\gamma_\mu B[\chi_-,u^\mu]\ra$&$M_1 B_1 \to M_2 B_2$\\
49&$\la\bar{B}[\chi_-,u^\mu]\gamma_\mu B\ra$&$M_1 B_1 \to M_2 B_2$\\
50&$\la\bar{B}u^\mu\ra\la\chi_-\gamma_\mu B\ra  
  -\la\bar{B}\chi_-\ra\la u^\mu\gamma_\mu B\ra$&$M_1 B_1 \to M_2 B_2$\\
51&$\la\bar{B}\{D_\mu f_+^{\mu\nu},\gamma_\nu B\}\ra$ & $B_1 \to \gamma B_2$\\
52&$\la\bar{B}[D_\mu f_+^{\mu\nu},\gamma_\nu B]\ra$& $B_1 \to \gamma B_2$\\
53&$i\la\bar{B}\gamma_5\gamma_\nu B[u_\mu,f_+^{\mu\nu}]\ra$&$\gamma
B_1 \to M_2 B_2$\\ 
54&$i\la\bar{B}[u_\mu,f_+^{\mu\nu}]\gamma_5\gamma_\nu B\ra$&$\gamma
B_1 \to M_2 B_2$\\
55&$i\left(\la\bar{B}u_\mu\ra\la f_+^{\mu\nu}\gamma_5\gamma_\nu
    B\ra - \la\bar{B}f_+^{\mu\nu}\ra\la u_\mu \gamma_5\gamma_\nu
    B\ra\right)$ &$\gamma B_1 \to M_2 B_2$\\ 
56&$\la\bar{B}\gamma^\tau
  B\{u^\mu,f_+^{\nu\rho}\}\ra\ve_{\mu\nu\rho\tau}$&$\gamma B_1 \to M_2 B_2$\\ 
57&$\la\bar{B}\{u^\mu,f_+^{\nu\rho}\}\gamma^\tau
  B\ra\ve_{\mu\nu\rho\tau}$&$\gamma B_1 \to M_2 B_2$\\ 
58&$\la\bar{B} u^\mu \gamma^\tau B
  f_+^{\nu\rho}\ra\ve_{\mu\nu\rho\tau}$ &$\gamma B_1 \to M_2 B_2$\\
59&$\la\bar{B}f_+^{\nu\rho}\gamma^\tau B u^\mu\ra
  \ve_{\mu\nu\rho\tau}$ &$\gamma B_1 \to M_2 B_2$\\
60&$\la\bar{B}\gamma^\tau B\ra\la  u^\mu
  f_+^{\nu\rho}\ra\ve_{\mu\nu\rho\tau}$&$\gamma B_1 \to M_2 B_2$\\
61&$\left(\la\bar{B}[u^\mu,f_+^{\nu\rho}]\sigma^{\lambda\tau}D_\mu
    B\ra - \la\bar{B}\overleftarrow{D}_\mu [u^\mu,f_+^{\nu\rho}]
\sigma^{\lambda\tau} B\ra\right)\ve_{\nu\rho\lambda\tau}$&$\gamma B_1 \to M_2 B_2$\\
62&$\left(\la\bar{B}\sigma^{\lambda\tau}D_\mu B [u^\mu,f_+^{\nu\rho}]\ra
- \la\bar{B}\overleftarrow{D}_\mu\sigma^{\lambda\tau} B
[u^\mu,f_+^{\nu\rho}]\ra\right)\ve_{\nu\rho\lambda\tau}$&$\gamma B_1 \to M_2 B_2$\\
63&$\left(\la\bar{B}u^\mu\ra\la f_+^{\nu\rho}\sigma^{\lambda\tau}D_\mu
    B\ra + \la\bar{B}\overleftarrow{D}_\mu f_+^{\nu\rho}\ra \la
    u^\mu\sigma^{\lambda\tau} B\ra\right)
  \ve_{\nu\rho\lambda\tau}$&$\gamma B_1 \to M_2 B_2$\\
64&$\la\bar{B}\{D_\mu f_-^{\mu\nu},\gamma_5\gamma_\nu B\}\ra$&$\gamma
B_1 \to M_2 B_2$\\ 
65&$\la\bar{B}[D_\mu f_-^{\mu\nu},\gamma_5\gamma_\nu B]\ra$&$\gamma
B_1 \to M_2 B_2$\\ 
66& $\la\bar{B}\gamma_5\gamma^\tau
  B\{u^\mu,f_-^{\nu\rho}\}\ra\ve_{\mu\nu\rho\tau}$&$\gamma B_1\to M_2
  M_3 B_2$\\
67&$\la\bar{B}\{u^\mu,f_-^{\nu\rho}\}\gamma_5\gamma^\tau
  B\ra\ve_{\mu\nu\rho\tau}$&$\gamma B_1\to M_2
  M_3 B_2$\\
68&$\la\bar{B}f_-^{\nu\rho}\gamma_5\gamma^\tau B u^\mu\ra
  \ve_{\mu\nu\rho\tau}$&$\gamma B_1\to M_2 M_3 B_2$\\
69&$\la\bar{B} u^\mu \gamma_5\gamma^\tau B
  f_-^{\nu\rho}\ra\ve_{\mu\nu\rho\tau}$&$\gamma B_1\to M_2 M_3 B_2$\\
70&$\la\bar{B}\gamma_5\gamma^\tau B\ra\la  u^\mu
  f_-^{\nu\rho}\ra\ve_{\mu\nu\rho\tau}$&$\gamma B_1\to M_2 M_3 B_2$\\
71&$i\la\bar{B}[u_\mu,f_-^{\mu\nu}]\gamma_\nu B\ra$&$\gamma B_1\to M_2
M_3 B_2$\\ 
72&$i\la\bar{B}\gamma_\nu B[u_\mu,f_-^{\mu\nu}]\ra$&$\gamma B_1\to M_2
M_3 B_2$\\ 
73&$i\left(\la\bar{B}u_\mu\ra\la f_-^{\mu\nu}\gamma_\nu B\ra
    -\la\bar{B}f_-^{\mu\nu}\ra\la u_\mu\gamma_\nu B\ra\right)$&$\gamma B_1\to M_2
M_3 B_2$\\ 
74&$i\left(\la\bar{B}\sigma_{\nu\rho} D_\mu B\{u^\mu,f_-^{\nu\rho}\}\ra
-\la\bar{B}\overleftarrow{D}_\mu\sigma_{\nu\rho}B\{u^\mu,f_-^{\nu\rho}\}\ra\right)$
&$\gamma B_1\to M_2 M_3 B_2$\\
75&$i\left(\la\bar{B}\{u^\mu,f_-^{\nu\rho}\}\sigma_{\nu\rho} D_\mu B\ra
-\la\bar{B}\overleftarrow{D}_\mu\{u^\mu,f_-^{\nu\rho}\}\sigma_{\nu\rho}B\ra\right)$
&$\gamma B_1\to M_2 M_3 B_2$\\
76& $i\left(\la\bar{B}u^\mu \sigma_{\nu\rho} D_\mu B f_-^{\nu\rho}\ra
-\la\bar{B}\overleftarrow{D}_\mu u^\mu \sigma_{\nu\rho} B
f_-^{\nu\rho}\ra\right)$
&$\gamma B_1\to M_2 M_3 B_2$\\
77&$i\left(\la\bar{B}f_-^{\nu\rho} \sigma_{\nu\rho} D_\mu B u^\mu \ra
-\la\bar{B}\overleftarrow{D}_\mu f_-^{\nu\rho}\sigma_{\nu\rho} B
u^\mu \ra\right)$ &$\gamma B_1\to M_2 M_3 B_2$\\
78&$i\left(\la\bar{B}\sigma_{\nu\rho} D_\mu B\ra - 
\la\bar{B}\overleftarrow{D}_\mu\sigma_{\nu\rho}B\ra\right)
\la u^\mu f_-^{\nu\rho}\ra $&$\gamma B_1\to M_2 M_3 B_2$\\
\hline
\end{longtable}

In the previous list, the symbol 
$D_{\nu\rho}=D_\nu D_\rho+D_\rho D_\nu$. For the other symbols 
we refer to \cite{OVP06}. In addition, a covariant derivative 
acts only on one hadronic matrix field, the one immediately  
to the right or 
left (in the latter case there is a left pointing arrow over $D$). 
E.g., $D_\rho B u_\nu$ must be understood 
such that the covariant derivative acts only on $B$.
 We also want 
to remark that our way of presenting the monomials of 
the ${\cal O}(q^3)$ meson-baryon chiral Lagrangian
 here and in \cite{OVP06} is much more compact and easy to manipulate 
than the one employed in \cite{FM06}. We also prefer
not to introduce dimensionful parameters 
to change artificially the dimension of the coefficients $h_i$. 
 
  In addition, we notice that
the monomial $O^{(3)}_{40}$ of \cite{FM06} is not exactly 
charge conjugation  invariant since 
those terms involving two covariant derivatives acting on the 
mesonic fields $u_\alpha$ are 
missed. These contributions, though ${\cal O}(q^5)$, are 
 needed to guarantee exact charge conjugation invariance. 

  It is  pointed out that 
the number of independent monomials in \cite{OVP06}
can be reduced by eight and that two
monomials were wrongly discarded in \cite{OVP06}. 
Then, a full  agreement in the number of independent
terms with \cite{FM06} arises.

\section*{Acknowledgements}

 We want to thank Matthias Frink for an exchange of emails. 

\end{document}